\documentclass[twocolumn,aps,superscriptaddress,nofootinbib]{revtex4-1}
\usepackage[dvips]{graphicx}
\usepackage{latexsym,amssymb,amsmath}
\usepackage{color}
\usepackage{enumerate}
\usepackage[bookmarksnumbered,bookmarksopen,colorlinks,citecolor=red,linkcolor=blue]{hyperref}
\usepackage{mathrsfs}
\usepackage{times}
\usepackage{csquotes}
\usepackage{multirow}
\usepackage[dvipsnames]{xcolor}
\usepackage{setspace}
\begin{document}

\title{Reconstructing the neutron star equation of state from observational data \\via automatic differentiation}

\author{Shriya Soma}
\email{soma@fias.uni-frankfurt.de}
\affiliation{Frankfurt Institute for Advanced Studies (FIAS), D-60438 Frankfurt am Main, Germany}
\affiliation{Institute f\"ur Theoretische Physik, Goethe Universit\"at, D-60438 Frankfurt am Main, Germany}
\affiliation{Xidian-FIAS International Joint Research Center, Giersch Science Center, D-60438 Frankfurt am Main, Germany}

\author{Lingxiao Wang}
\email{lwang@fias.uni-frankfurt.de}
\affiliation{Frankfurt Institute for Advanced Studies (FIAS), D-60438 Frankfurt am Main, Germany}
\affiliation{Xidian-FIAS International Joint Research Center, Giersch Science Center, D-60438 Frankfurt am Main, Germany}

\author{Shuzhe Shi}
\email{shuzhe.shi@stonybrook.edu}
\affiliation{Center for Nuclear Theory, Department of Physics and Astronomy, Stony Brook University, Stony Brook, New York, 11794, USA.}

\author{Horst St\"ocker}
\email{stoecker@fias.uni-frankfurt.de}
\affiliation{Frankfurt Institute for Advanced Studies (FIAS), D-60438 Frankfurt am Main, Germany}
\affiliation{Institute f\"ur Theoretische Physik, Goethe Universit\"at, D-60438 Frankfurt am Main, Germany}
\affiliation{GSI Helmholtzzentrum f\"ur Schwerionenforschung GmbH, D-64291 Darmstadt, Germany}

\author{Kai Zhou}
\email[Corresponding author : ]{zhou@fias.uni-frankfurt.de}
\affiliation{Frankfurt Institute for Advanced Studies (FIAS), D-60438 Frankfurt am Main, Germany}

\date{\today}

\begin{abstract}
%The equation of state (EoS) that describes strong interactions in extremely dense matter remains mysterious. 
%The equation of state~(EoS) of dense matter with strong interactions is not well understood. First-principle lattice quantum chromodynamics~(LQCD) calculations of the EoS at large chemical potential are challenging. However, neutron star observables like masses, radii, moments of inertia and tidal deformability are direct probes to the EoS. Hence, such data make the EoS reconstruction task feasible. In this work, results from a novel deep learning technique that optimize a parameterized equation of state in the automatic differentiation~(AD) framework of solving inverse problems are presented. Stellar structures are predicted from a pre-trained Tolman–Oppenheimer–Volkoff solver network, given an EoS represented by neural networks. Observational data of neutron stars, specifically their masses and radii, are used to implement the chi-square fitting. The parameters of the neural network EoS are optimized by minimizing the error between observations and predictions. The well-trained neural network EoS yields rather narrow bands for the relationship between the pressure and speed of sound as a function of the mass density. The results presented are consistent with those obtained from conventional approaches and the experimental bound on the tidal deformability inferred from the gravitational wave event, GW170817. 
Neutron star observables like masses, radii, and tidal deformability are direct probes to the dense matter equation of state~(EoS). A novel deep learning method that optimizes an EoS in the automatic differentiation framework of solving inverse problems is presented. The trained neural network EoS yields narrow bands for the relationship between the pressure and speed of sound as a function of the mass density. The results are consistent with those obtained from conventional approaches and the observational bound on the tidal deformability inferred from the gravitational wave event, GW170817. 
\end{abstract}

\maketitle

\emph{Introduction.  }Neutron stars~(NSs) serve as cosmic laboratories for the study of neutron-rich nuclear matter~\cite{baym:2018hadrons}, with central densities far greater than the nuclear saturation density~($\rho_0=0.16~\text{fm}^{-3}$). Heavy-ion collisions provide means to compress nuclear matter in terrestrial laboratories to such high densities, but inevitably involve moderate to high temperatures~\cite{Stoecker:1986ci,  xu:2021littlebang, Dexheimer:2020zzs}. Non-perturbative lattice quantum chromodynamics~(LQCD) calculations for the finite-density region of the QCD phase diagram are a long-standing theoretical challenge due to the sign-problem~\cite{aarts:2016introductory}. Nevertheless, studies dedicated to probe the cold and dense nuclear matter properties have benefited from astronomical observations in past decades~\cite{watts:2016colloquium}. NS observables, e.g., mass, radius, and tidal deformability are highly dependent on the underlying dense matter bulk properties -- the nuclear equation of state (EoS)~\cite{kojo:2021qcd}. With the increasing number of observations of long-lived single NSs and colliding NSs, as of the operation of gravitational wave~(GW) detectors and associated multi-messenger astrophysics, the inference of the dense QCD matter EoS appears promising~\cite{Hanauske:2017oxo, Most:2019onn, Huth:2021bsp, Yunes:2022ldq}.

The EoS at low densities~($\rho \lesssim 1-2 \rho_0$) can be extrapolated from finite nuclei experiments and can be calculated from the chiral effective field theory~($\chi$EFT)~\cite{Hebeler:2013nza,Tews:2012fj,Drischler:2017wtt}. At extremely high densities~(~$\gtrsim~40\rho_0$), one may resort to perturbative QCD~(e.g., the hard dense loop) calculations~\cite{kurkela:2010cold,Komoltsev:2021jzg}. The intermediate density regime~(~$\sim~2-10 \rho_0$) is however not accessible to QCD. This high baryon density region is usually calculated from different effective field theoretical models~\cite{oertel:2017equations}, e.g., pure nucleonic EoSs~\cite{Sugahara:1993wz}, hybrid (hadrons and quarks) EoSs~\cite{Motornenko:2019arp}, and hyperonic EoSs~\cite{Banik:2014qja}. %Tolman--Oppenheimer--Volkoff~(TOV) equations~\cite{PhysRev.55.364, PhysRev.55.374} are routinely solved to obtain the corresponding structural properties of static NSs for a given EoS.
The existence of a one-to-one mapping from the mass-radius ($M$-$R$) curve of NSs to the corresponding EoS~\cite{lindblom:1992determining} provides possibilities to reconstruct the EoS model-independently via an inverse process. However, the limited number of NS observations and their large measurement uncertainties pose severe difficulties in accurately inferring the underlying EoS.

As a modern computational paradigm, deep learning is tailored to represent indirect mappings or find hidden structures in complex systems. It has been effective in solving a number of physics problems, such as determining the parton distribution function~\cite{Forte_2002,Collaboration_2007,Gao:2022iex}, reconstructing the spectral function~\cite{Kades:2019wtd, Zhou:2021bvw, Chen:2021giw}, and identifying phase transitions and impact parameters for heavy-ion collisions~\cite{Pang:2016vdc,Du:2019civ, Wang:2020hji, Wang:2020tgb, OmanaKuttan:2020brq, OmanaKuttan:2020btb,Jiang:2021gsw}. Recent works implement machine learning in a supervised learning manner. These include, for example, the non-linear mapping between a mass-radius~($M$-$R$) curve and its corresponding EoS~\cite{fujimoto:2018methodology, fukushimaPhysRevD.101.054016, fujimoto:2021extensive,morawski:2020neurala,Ferreira_2021,galaxies10010016}. The learned mapping is represented by a neural network which predicts the EoS or its parameters from NS observational data. The supervised learning algorithms constrain the EoS in a reasonable range. % with
However, using a specific training data set in the data-hungry network may lead to generalization errors~\cite{goodfellow:2016deep}. 
In addition, the uncertainty estimate for EoS thus reconstructed lacks principled guidance. As an alternative, the EoS can be parameterized in a model-independent way, and the parameters inferred from finite observations. This can be achieved through Gaussian processes~\cite{essick:2020nonparametric,landry:2020nonparametric} or by the conventional Bayesian inference\cite{Steiner_2010, raithel:2017neutron, Traversi_2020,Thete:2022eif}. In the latter approach, one could make use of shallow neural networks to parameterize the EoS and determine its parameters in an unsupervised way~\cite{Han:2021kjx}. A flexible parameterization may, however, lead to a computationally expensive optimization process.
%%%%%%%%%%%%%%%%%%%%%%%%%%%%%%%%%%%%%%%%%%%%%%%%%%%%%%%%
\begin{figure*}
\begin{center}
\includegraphics[width=0.85\textwidth]{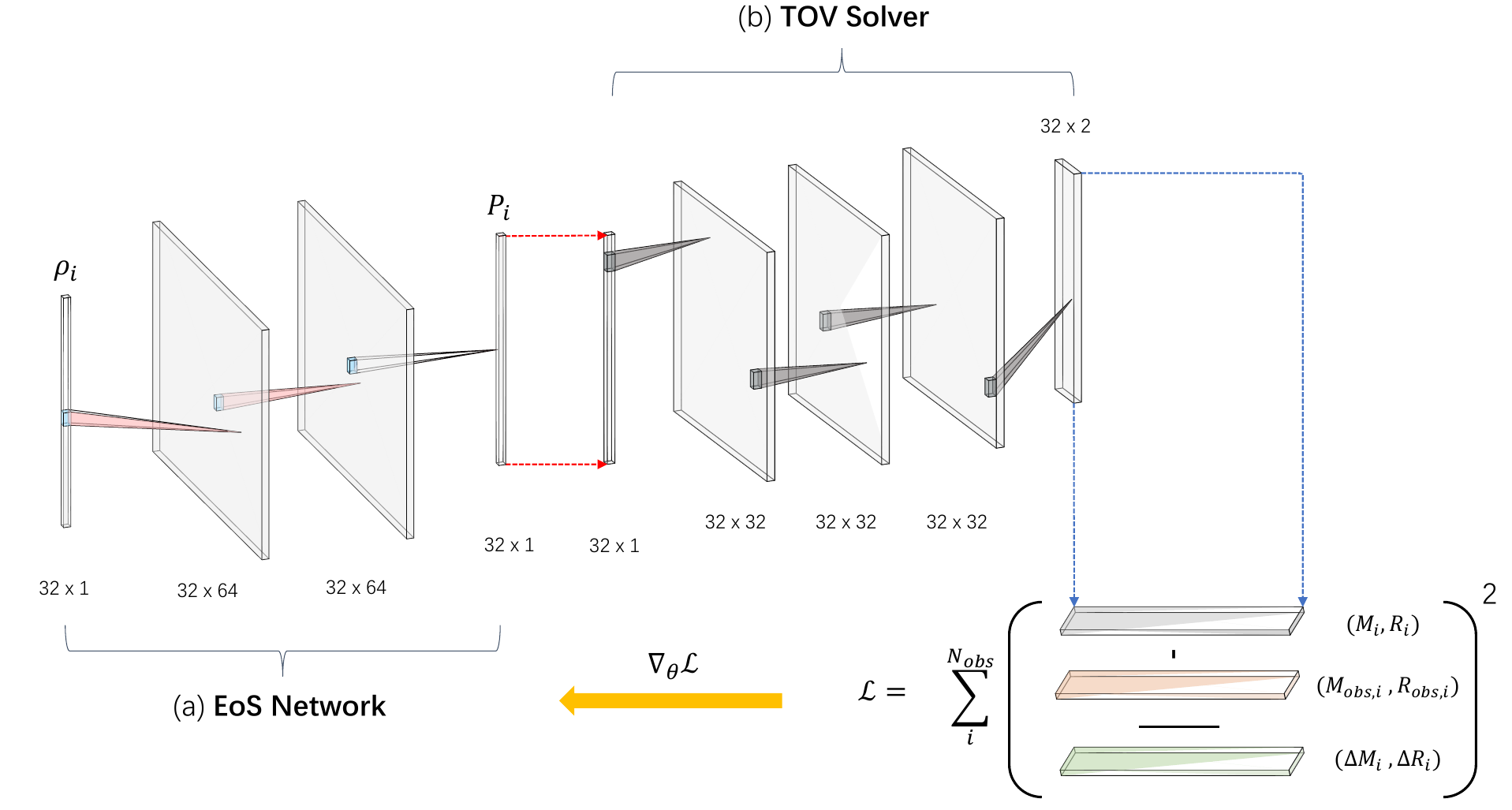}
\caption{A flow chart of the developed method developed. The \texttt{EoS Network}~(a) is a neural network representation of the EoS. The \texttt{TOV-Solver}~(b) is a well-trained static network.}\label{fig:framework}
\end{center}
\end{figure*}
%%%%%%%%%%%%%%%%%%%%%%%%%%%%%%%%%%%%%%%%%%%%%%%%%%%%%%%%%%%
It has been shown that physics-driven deep learning tools can surpass traditional methods in solving {\it inverse problems}~\cite{Shi:2021qri, Wang:2021jou}.
To reconstruct the NS EoS in an unbiased manner, a novel approach which utilizes deep neural networks~(DNNs) in the automatic differentiation~(AD) framework is introduced. This method has been tested on mock data in our previous publication~\cite{Soma:2022qnv}. In the present work, the NS EoS is studied based on the same approach, by utilizing real observational NS data. The current experimental constraints from GW data and possible phase structures coded in EoS are discussed.

\emph{Methods.  }\label{sec:methods}
% \subsection{Reconstructing EoS via automatic differentiation} \label{sub:ad}
The AD framework developed here shown in Fig.~\ref{fig:framework} consists of two differentiable modules: the \texttt{EoS Network}, $P_{\theta}(\mathbf{\rho})$, an unbiased and flexible parameterization of the EoS using DNNs; and the \texttt{TOV-Solver}, a DNN for translating any given EoS to its corresponding $M$-$R$ curve. The latter is an emulator for solving the Tolman--Oppenheimer--Volkoff~(TOV) equation~\cite{PhysRev.55.364, PhysRev.55.374} which is solved to obtain the structural properties of non-rotating NSs.

The \texttt{EoS Network}, when combined with the well-trained \texttt{TOV-Solver} network, can be optimized in an unsupervised manner. Given $N_{\text{obs}}$ number of NS observations, the \texttt{EoS Network} is trained to fit the predictions of pairwise ($M$, $R$) from the pipeline (\texttt{EoS Network} $+$ \texttt{TOV-Solver}) to observations. A gradient-based algorithm within AD framework is deployed to minimize the loss function, $\chi^2$, expressed as
\begin{equation}
\chi^2 = \sum_{i=1}^{N_{\text{obs}}} \frac{(M_{i} - M_{\text{obs},i})^2}{\Delta M_i^2}
+    \frac{(R_{i} - R_{\text{obs},i})^2}{\Delta R_i^2}. \label{eq:chi2}
\end{equation}
Here ($M_{i},R_i$) represents the output of the \texttt{TOV-Solver}, and ($M_{\text{obs},i},R_{\text{obs},i}$) are observations which have an uncertainty ($\Delta M_{i},\Delta R_i$). With a static, well-trained \texttt{TOV-Solver} network, the gradients of the loss with respect to parameters of the \texttt{EoS Network} are
\begin{equation}
\frac{\partial\chi^2}{\partial \theta} = \sum_{i=1}^{N_{\text{obs}}}\int
\bigg[\frac{\partial\chi^2}{\partial M_i} \frac{\delta M_i}{\delta P_{\theta}(\rho)}
+\frac{\partial\chi^2}{\partial R_i} \frac{\delta R_i}{\delta P_{\theta}(\rho)}\bigg]
\frac{\partial P_{\theta}(\rho)}{\partial \theta} \mathrm{d}\rho.
\label{eq:ad}
\end{equation}
%\begin{equation}
%\frac{\delta\chi^2}{\delta \theta} = \sum_{i=1}^{N_{\text{obs}}}\frac{\delta\chi^2}{\delta (M_i, R_i)} \frac{\delta (M_i, R_i)}{\delta P_{\theta}(\rho)}\frac{\delta P_{\theta}(\rho)}{\delta \theta},\label{eq:ad}
%\end{equation}
%\begin{equation}
%\frac{\delta\chi^2}{\delta \theta} = \frac{\delta\chi^2}{\delta \mathbf{z}} \frac{\delta \mathbf{z}}{\delta x}\frac{\delta x}{\delta \theta},\label{eq:ad}
%\end{equation}
%where $\mathbf{z}=(M_i,R_i)$, $x= P_j(\rho_i\j)$ and 
Here, the \texttt{TOV-Solver} is a mapping $f: P_{\theta}(\rho) \rightarrow {(M_i, R_i)}$.
%\ssz{Why do we denote vector $z$ in bold font but not $x$? Also, use subscript $i$ to label both $M$ and $\rho$ might confuse people? maybe $j$ for latter?} 
The last two terms, i.e., the partial derivative ${\partial P_{\theta}}/{\partial \theta}$ and the functional derivative ${\delta (M_i, R_i)}/{\delta P_{\theta}(\rho)}$, can be directly computed via back-propagation algorithm~\cite{goodfellow:2016deep} within the AD framework for the two coupled DNNs. The network is composed of a series of differentiable modules including linear transformations and nonlinear activation functions. The details are shown in Fig.~\ref{fig:framework} and explanations are given in following sections. The parameters of the \texttt{EoS Network} are optimized to obtain the best fit from the well-trained \texttt{TOV-Solver} to the finite and noisy observational $M$-$R$ data. We implement the above framework using the Tensorflow~\cite{tensorflow2015-whitepaper} library.

% \subsection{EoS Network} \label{eos-network}

The \texttt{EoS Network} is structured with baryon number density~($\rho=[1,7.4]\rho_0$) as the input, and the corresponding pressure~($P(\rho)$) as the output. The input density is chosen to be a linearly spaced 1D array of length $N_{\rho}$= 32 on a logarithmic scale, normalized to lie within the range $(0,1)$. The input shape of the network is (height,channels)=$(32,1)$ as shown in Fig.~\ref{fig:framework}, with the height representing the number of discrete density points, and the channel indicating the number of trainable kernels in subsequent layers. The 1D convolutional kernel with non-negative activation function introduces a monotonicity in the mapping between density and pressure. This is due to the shared parameters along each density value across the height dimension. This ensures thermodynamic stability for the reconstructed EoS. The architecture of the \texttt{EoS Network} is shown in Fig.~\ref{fig:framework}(a), where two hidden layers link the input and output with activation function, ELU. The output ($P_{\theta}(\rho)$) from the \texttt{EoS Network} is further transmitted to the well-trained \texttt{TOV-Solver}, which returns a prediction of the $M$-$R$ pairs.

The training of the \texttt{EoS Network} is in an unsupervised learning framework, proceeded by maximizing the likelihood function of the observational data. The trainable parameters are the weights and biases of the networks. The optimization procedure involves minimizing the $\chi^2$ loss shown in Eq.~\eqref{eq:chi2}. There are recent works that parameterize the EoS with shallow neural networks or Gaussian processes~\cite{Han:2021kjx, Legred:2022pyp}. The method proposed in this work realizes the deep (instead of shallow) neural network as a flexible representation of the EoS with an efficient and economic optimization.%, which is challenging for previous works.

% \subsection{TOV-Solver} \label{tov}
The TOV equation is given as
%%%%%%%%%%%%%%%%%%%%%%%%%%%%%%%%%%%%%%%%%%%%%%%%%%
\begin{equation}
    -\frac{dP}{dr} = \frac{\big{[}\epsilon(r)+P(r)\big{]}\big{[}m(r)+4\pi r^3 P(r)\big{]}}{r[r-2m(r)]}\,,
\end{equation}
%%%%%%%%%%%%%%%%%%%%%%%%%%%%%%%%%%%%%%%%%%%%%%%%%%
and
%%%%%%%%%%%%%%%%%%%%%%%%%%%%%%%%%%%%%%%%%%%%%%%%%%
\begin{equation}
\frac{dm(r)}{dr} = 4\pi r^2 \epsilon (r) ,
\end{equation}
%%%%%%%%%%%%%%%%%%%%%%%%%%%%%%%%%%%%%%%%%%%%%%%%%%%
where $r$ is the radial distance from the centre of the star, and $m(r)$ is the mass enclosed within $r$. We use the natural unit ($c=G=1$) throughout this letter. In order to determine the corresponding observables given an EoS, i.e, mass ($M$) and radius ($R$) of the star, the TOV equations are integrated radially outwards from the centre. The initial conditions are $P(r=0)=P_c$, where $P_c$ is the central pressure. The radius of the star, $R$, is defined from the boundary condition on the surface, $P(r=R)=0$, and the mass enclosed in $R$, i.e, $M=m(R)$, is the total mass of the star. 

Training data is prepared using a large set of EoSs and their corresponding $M$-$R$ sequences to obtain an effective \texttt{TOV-Solver} represented by DNNs. Four different low density~($\rho<\rho_0$) EoSs are included in the training data generation: PS, SLy, DD2 and TM1. The 2$M_{\odot}$ constraint obtained from observational data is applied~\cite{Demorest2010, antoniadis:2013massive, Fonseca_2021, Romani:2021xmb}. A less conservative bound is used in the training data, i.e, the $M$-$R$ sequences of the EoS must accommodate a neutron star of mass 1.9$M_{\odot}$.
%corresponding to PS~\cite{}; 61,273 EoSs corresponding to SLy~\cite{}; and 72,834 EoSs corresponding to DD2~\cite{}. We first derive the mass-radius ($M$-$R$) sequences for all the EoSs and the left must meet the criteria of accommodating a neutron star of mass 1.9$M_{\odot}$, a constraint obtained from observational data~\cite{Demorest2010, antoniadis:2013massive, cromartie:2020relativistic}.
The details of training and validation can be found in our preliminary work~\cite{Soma:2022qnv}. In summary, the well-trained network can efficiently obtain the $M$-$R$ pairs given an arbitrary EoS, and one typical architecture is exhibited in Fig~\ref{fig:framework}(b), in which the fundamental modules are consistent with the above descriptions but they are non-trainable.

%We therefore set EoSs as the input to the network and train it to output their corresponding $M$-$R$ sequences.
% \subsection{Processing observational data}\label{sub:data}
Normal distributions are used to fit the observational data in a similar procedure as in Ref~\cite{fukushimaPhysRevD.101.054016,fujimoto:2021extensive}. The data is available in the form of posterior probability distributions~\cite{Ozel:2015fia, Nattila:2017wtj, Gonzalez-Caniulef:2019wzi, riley:2019nicer, Riley_2021}. The observations are characterized with the means and the variances of the fitted marginal distributions of $M$ and $R$. This is also in accordance with the loss function used in the current methodology as shown in Eq.~\eqref{eq:chi2}.
%Table~\ref{tab:obs} shows the fitted contour for (arbitrarily chosen) one representative out of 14 observations.
Table. I in the Appendix shows the fitted values of $M$ and $R$ for the current $M$-$R$ relation on NSs determined by various astrophysical observations.

% \subsection{Uncertainty Estimation}
\emph{Uncertainty Estimation. }
The Bayesian perspective is invoked to estimate the uncertainties of the reconstructed EoS from the method proposed here.
%In this section we discuss the manifestation of uncertainty from the $M$-$R$ observations in the reconstructed EoS. In a realistic scenario of reconstructing the EoS from observational data, uncertainty is naturally accompanied. 
The training loss function is defined as the distance square between the observed and predicted $M$-$R$ pairs in our method, which is the negative log-likelihood for given observations. However, the observed data is not uniformly distributed on the $M$-$R$ plane, and the measurement errors tend to cause a disarrangement in the $M$-$R$ relationship. Furthermore, the data is limited and the corresponding central densities are unknown.
%{In such circumstances, a finer resolution ($N_{\rho}=128$) might prove useful.}
To tackle this ambiguity in the sense of maximizing likelihood, we adopt the method of ``closest approach'' ~\cite{raithel:2017neutron} in the $\chi^2$ loss, for each iteration of the training, i.e. 
\begin{equation}
    \chi^2(M,R) = \sum_{i=1}^{N_{\text{obs}}} \frac{(M(\rho_{ci}) - M_{\text{obs},i})^2}{\Delta M_i^2}
+    \frac{(R(\rho_{ci}) - R_{\text{obs},i})^2}{\Delta R_i^2}\,.
\label{eq:loss}
\end{equation}
$\rho_{ci}$ is updated for every $i^\text{th}$ observation as,
\begin{equation}
    \rho_{ci} =\arg \min_{\rho_c} \frac{(M(\rho_{c}) - M_{\text{obs},i})^2}{\Delta M_i^2}
+    \frac{(R(\rho_{c}) - R_{\text{obs},i})^2}{\Delta R_i^2}\,.
\label{eq:update_ind}
\end{equation}
Eq.~\eqref{eq:update_ind} is therefore used to determine the central densities of the star ($M_{\text{obs}},R_{\text{obs}}$) that result in the least distance between the predicted and the real $M$-$R$ pairs. Using this method, the measurement uncertainties of observational data is naturally encoded in the proposed reconstruction method to obtain the corresponding posterior distribution of EoS, from which the uncertainty of the resulting EoS is obtained.
The details of calculating the uncertainty can be found in the Appendix. The causal condition (speed of sound $c_s^2<1$) is also introduced to eliminate non-realistic reconstructions.  

%%%%%%%%%%%%%%%%%%%%%%%%%%%%%%%%%%%%%%%%%%%%%%%%%%%%%%%%%%%%
\begin{figure}[htbp!]
\centering
\includegraphics[width=0.95\columnwidth]{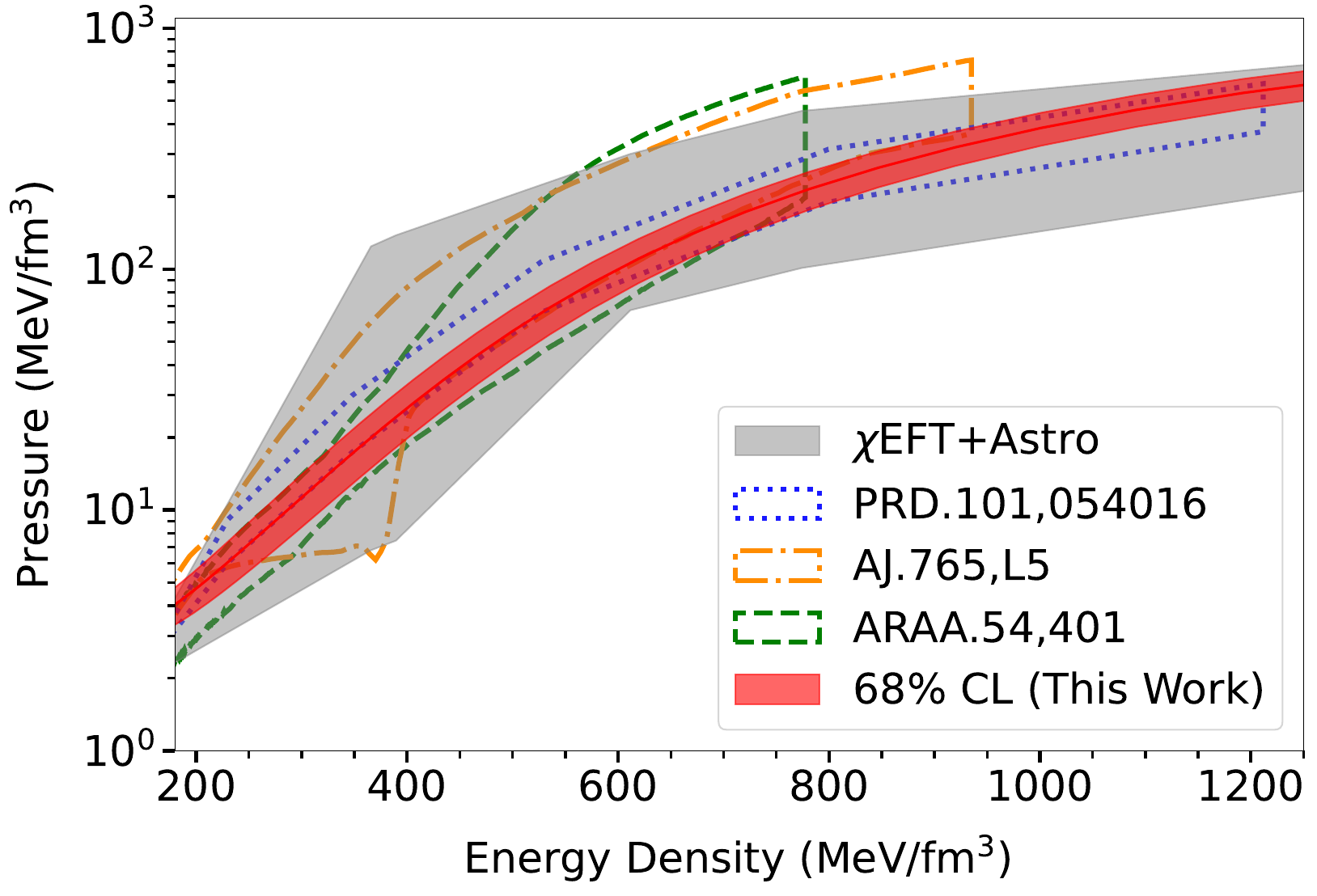}
\caption{The EoS reconstructed from observational data of 18 neutron stars (labelled as ``This Work''). The red shaded area represents the 68\% confidence interval~(CI) evaluated directly from reconstruction. Other constraints on the EoS like the $\chi$EFT prediction (gray band), results derived from Bayesian methods (AJ.765,L5~\cite{steiner:2013neutron} and ARAA.54,401~\cite{ozel:2016masses}) and the direct inverse mapping (PRD.101,054016~\cite{fukushimaPhysRevD.101.054016}) are also included.}\label{fig:eos}
\end{figure}
%%%%%%%%%%%%%%%%%%%%%%%%%%%%%%%%%%%%%%%%%%%%%%%%%%%%%%%%%%%%
\emph{Results and Discussion.  }\label{sev:results}
The dense matter EoS is reconstructed from the proposed method using the 18 chosen NS mass-radius observations. Fig.~\ref{fig:eos}, presents the resulting EoS within 1$\sigma$ confidence interval~(CI) as a light red shaded band~(labelled ``This Work"). In order to reconstruct the EoS, 10,000 $M$-$R$ curves are sampled based on Table. I in Appendix. Post a causality screening which rejects all reconstructed EoSs with superluminous speeds of sound, the uncertainty estimates are extracted for the approach described above. %in Sec.\ref{sec:methods}.
For comparison, the results of other works are also shown in the figure. The results from the method proposed here lie within the limits estimated from chiral effective theory calculations ($\chi$EFT). The band of uncertainties is considerably narrower than other Bayesian analyses and recent supervised learning predictions. The band thus reconstructed is a smooth curve which is due to the flexible neural network representation. The DD2 EoS is adopted for sub-saturation densities. While low-energy nuclear experimental data help constrain the reconstructed EoS in this density region~\cite{Margueron:2017eqc,Margueron:2017lup}, the EoS at intermediate density regions can be probed from the existing NS observations. The reconstruction method used here focuses on the most uncertain high density region ($\rho\approx~$2-8$\rho_0$), and can therefore be used to deduce the EoS from an ever increasing number of observations.

%%%%%%%%%%%%%%%%%%%%%%%%%%%%%%%%%%%%%%%%%%%%%%%%%%%%%%%%%%%%
\begin{figure}[htbp!]
\centering
\includegraphics[width=0.95\columnwidth]{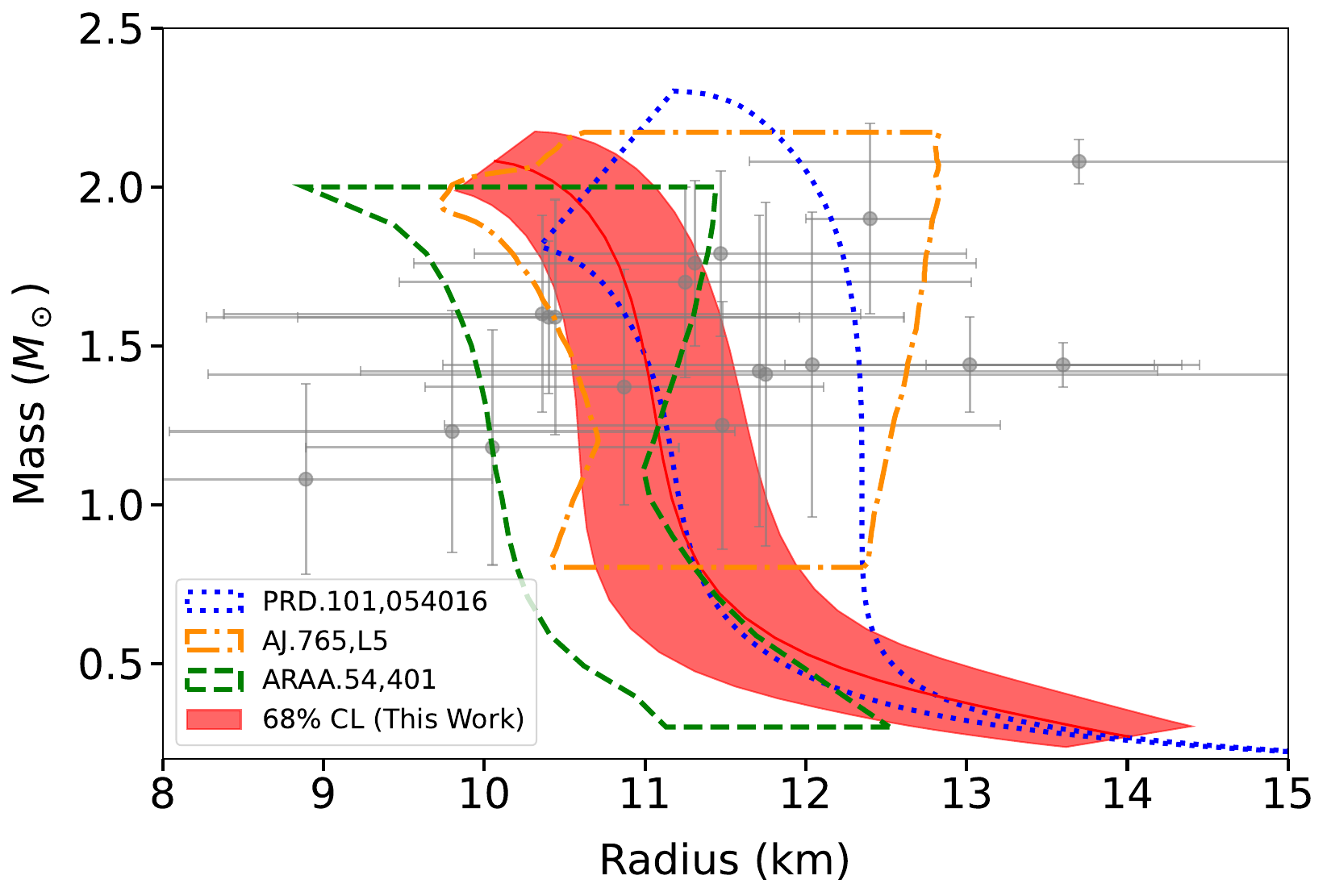}
\caption{$M$-$R$ contour corresponding to the reconstructed EoS (This Work) in Fig.2 and the other EoS candidates. The grey dots with uncertainties are our fitting observations summarized in Table. 1 in the Appendix.}\label{fig:mr}
\end{figure}
%%%%%%%%%%%%%%%%%%%%%%%%%%%%%%%%%%%%%%%%%%%%%%%%%%%%%%%%%%%%

The $M$-$R$ bands and contours which correspond to the reconstructed EoSs in Fig.~\ref{fig:eos} are shown in Fig.~\ref{fig:mr}. Astrophysical data (Gaussian fitted as shown in Appendix.) %Appendix.~\ref{app}) 
used in this study are also plotted. The observational ``massive neutron star" constraint has also been implemented, i.e, reconstructed $M$-$R$ curves with maximum mass $M_\text{max}\leq 1.9 M_\odot$ are excluded. Our results exhibit a narrower band as compared to other works. This is partly also due to the low density EoS prior chosen in the current analysis. The reconstructed EoS (red band) certainly supports massive neutron stars ($>2M_{\odot}$), and predicts $R_{1.4}= 11.1\pm 0.51\,$km (at $68\%$ confidence level (CL)) for a canonical $1.4 M_{\odot}$ neutron star, which is consistent with recent constraints from multi-messenger observations~\cite{Capano:2019eae}.

%%%%%%%%%%%%%%%%%%%%%%%%%%%%%%%%%%%%%%%%%%%%%%%%%%%%%%%%%%%%
\begin{figure}[htbp!]
\centering
\includegraphics[width=0.95\columnwidth]{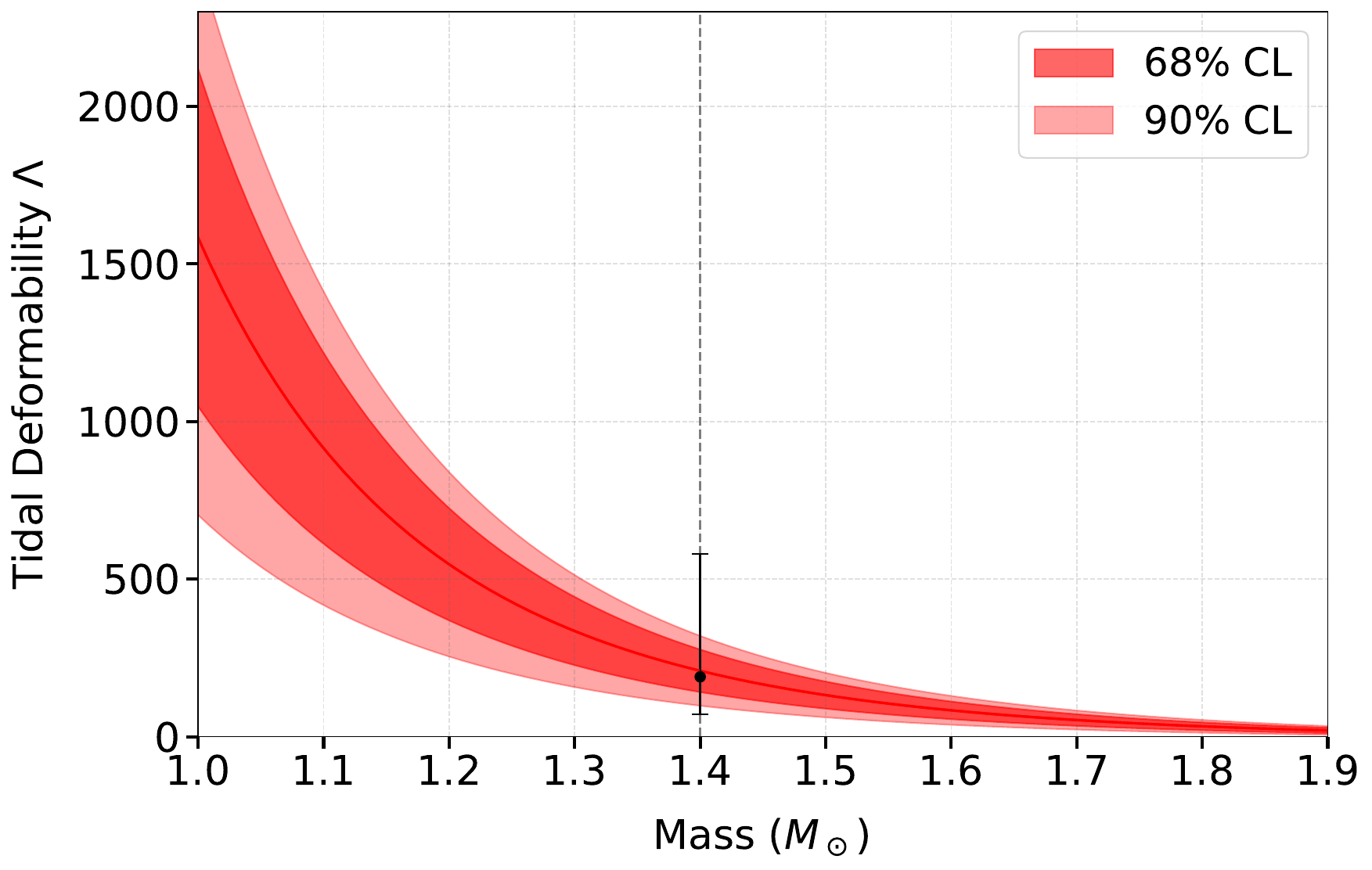}
\caption{Tidal deformability derived from the reconstructed EoS. The $90\%$ CL of the GW observation, GW170817, is shown as black bar.}\label{fig:td}
\end{figure}
%%%%%%%%%%%%%%%%%%%%%%%%%%%%%%%%%%%%%%%%%%%%%%%%%%%%%%%%%%%%
Further validation of the proposed method can be performed directly by confronting the {\it predicted} tidal deformability, $\Lambda$, from the reconstruction, with the constraints on $\Lambda$ as obtained from the gravitational wave
%We further validate the proposed method and the results obtained from the limits on tidal deformability, $\Lambda$, obtained from the recent gravitational wave 
observation, GW170817~\cite{Annala:2017llu,LIGOPhysRevLett.121.161101}.
From the reconstructed EoSs, the corresponding $\Lambda$ is computed as a function of NS mass, as shown in Fig.~\ref{fig:td}. Here, the recent constraint from the event GW170817, $\Tilde{\Lambda}_{1.4}=190_{-120}^{+390}$ \cite{LIGOPhysRevLett.121.161101}(at $90\%$ CL), is also shown. The value obtained from the reconstructed EoSs in this work, $\Lambda_{1.4}=209.12_{-110.8}^{+110.8}$ (at $90\%$ CL) is in good agreement with the bounds estimated from GW170817.

%%%%%%%%%%%%%%%%%%%%%%%%%%%%%%%%%%%%%%%%%%%%%%%%%%%%%%%%%%%%
\begin{figure}[htbp!]
\centering
\includegraphics[width=0.95\columnwidth]{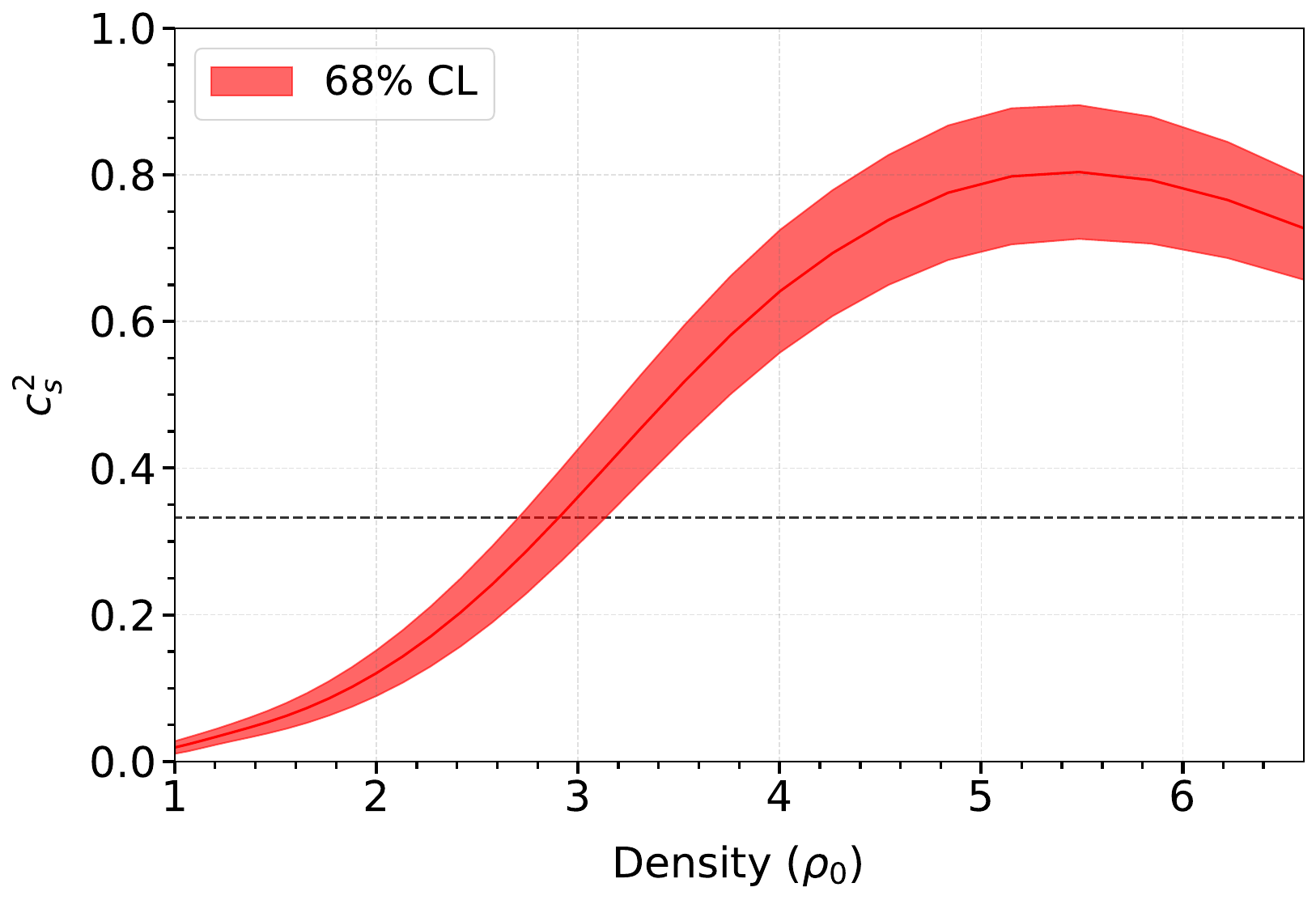}
\caption{Speed of sound corresponding to the reconstructed EoS in Fig.2. The band represents 68\% CL. The horizontal dotted line is the conformal limit of $c_s^2 = 1/3$.}\label{fig:sound}
\end{figure}
%%%%%%%%%%%%%%%%%%%%%%%%%%%%%%%%%%%%%%%%%%%%%%%%%%%%%%%%%%%%

The speed of sound in the neutron star matter is another important characteristic of the EoS. The squared value,~$c_s^2$ is derivable from any given EoS as $c_s^2 = \partial P/\partial \epsilon$. Fig.~\ref{fig:sound} displays the corresponding $c_s^2$ of the EoS reconstructed in this work, with natural unit $c = 1$. At low and medium densities ($\rho<3\rho_0$), $c_s^2$ shows an increase up to 0.3, with a relatively narrow band of uncertainty. It does exceed the conformal limit (the limits for massless non-interacting ultra-relativistic matter) in the high-density region ($\rho>3\rho_0$), which implies the presence of strong interactions in the dense matter. $c_s^2$ seems to saturate or decrease beyond $\rho\sim5-6\rho_0$. A smooth decrease of $c^2_s$ might eventually reach down to the conformal limit, found in recent calculations~\cite{Tews:2018kmu,fukushimaPhysRevD.101.054016} which interpret this as the approach of the asymptotically free state of quarks and gluons in a hadron-quark continuity picture~\cite{Fukushima:2015bda,Kojo:2020krb}. However, due to limited observations and their huge uncertainties, the possibility of a phase transition cannot be ruled out. If the number of observations and certainty can both be improved significantly in the near future, the possible existence and the order of phase transitions can be recognized in the present framework. This has been validated on a mock data-set with sufficiently precise mass-radius pairs.

\emph{Summary.  }\label{summary}This work implements a novel AD-based approach with DNNs for {\it inverse problem} solving, namely, to reconstruct the NS EoS using $M$-$R$ data from 18 NS observations. Under a Bayesian inference picture, the EoS is represented by DNNs (\texttt{EoS Network}) without explicit physical model priors. A combination with a well-trained (\texttt{TOV-Solver}) network for emulating the TOV equations solver, yields an AD which is naturally applied to further perform the EoS reconstruction with modern gradient-based optimizations. Such an unsupervised learning manner allows for reconstructing the underlying EoS with uncertainties corresponding to the observations. The causality limit and the massive mass observation are adopted as post-constraints. A further check based on current gravitational wave observations shows remarkable consistency with the existing tidal deformability prediction. The physics-driven deep learning approach established here achieves a successful reconstruction of the NS EoS. The results are in agreement with current data. Multi-messenger observations can be incorporated into the proposed method in the near future.

\textbf{Acknowledgment. } The authors thank Dr. Yuki Fujimoto, Kenji Fukushima, Zidu Lin, Anton Motornenko, and Jan Steinheimer for useful discussions. The work is supported by (i) Deutscher Akademischer Austauschdienst - DAAD (S. Soma) (ii) the BMBF under the ErUM-Data project (K. Zhou), (iii) the AI grant of SAMSON AG, Frankfurt (S. Soma, L.Wang and K. Zhou), (iv) Xidian-FIAS International Joint Research Center, ECT$^*$ workshop supported by Strong 2020, GSI/EMMI and INFN (L. Wang), (v) U.S. Department of Energy, Office of Science, Office of Nuclear Physics, grants Nos. DE-FG88ER40388 and DE-SC0012704 (S. Shi), (vi) by the Walter Greiner Gesellschaft zur F\"orderung der physikalischen Grundlagenforschung e.V. through the Judah M. Eisenberg Laureatus Chair at Goethe Universit\"at Frankfurt am Main (H. St\"ocker). We also thank the donation of NVIDIA GPUs from NVIDIA Corporation.

%\ssz{update [44,45] (upon publication) }

\bibliography{nnEoS}
\bibliographystyle{apsrev4-1}
%\newpage

\appendix

\section{Data preprocessing and importance sampling}\label{app}
To evaluate the uncertainty of the reconstruction, one can adopt Bayesian perspective to focus on the posterior distribution of the EoSs for the given astrophysical observations, $\text{Posterior}(\boldsymbol{\theta}_{\text{EoS}}|\text{data})$. In this calculation, we first draw an ensemble of $M-R$ samples from the fitted Gaussian distribution of real observations (in Table~\ref{tab:obs}). From this ensemble, %(to be the references)
we deterministically infer the corresponding EoS with maximum likelihood estimation. Given the ensemble of reconstructed EoSs, one can then apply importance sampling to estimate the uncertainty related to the desired posterior distribution, where a proper weight is evaluated to each EoS.
Our results and uncertainty estimations in the main text all obey this strategy. In general, a physical variable $\hat{O}$ can be estimated as,
\begin{equation}
    \bar{O} = \langle \hat{O} \rangle = \sum_j^{N_\text{samples}} w^{(j)} O^{(j)}.
\end{equation}
The standard deviation is given by $(\Delta O)^2 = \langle \hat{O}^2 \rangle - \bar{O}$. The weights are,
\begin{align}
    w^{(j)} &= \frac{\text{Posterior}(\boldsymbol{\theta}^{(j)}_{\text{EoS}}|\text{data})}{\text{Proposal}(\boldsymbol{\theta}^{(j)}_{\text{EoS}})} \nonumber \\
      &= \frac{p(\text{data}|\boldsymbol{\theta}^{(j)}_{\text{EoS}})\text{Prior}(\boldsymbol{\theta}^{(j)}_{\text{EoS}})}{p(\boldsymbol{\theta}^{(j)}_{\text{EoS}}|\text{samples}^{(j)})p(\text{samples}^{(j)}|\text{data})},
\end{align}
where $j$ indicates the index of reconstructed EoS (from the several samples of EoS), $\boldsymbol{\theta}_{\text{EoS}}$ is the parameter set representing the EoS, $p(\text{samples}|\text{data})=\mathcal{N}(M_{\text{obs}},\Delta{M}^2)\mathcal{N}(R_{\text{obs}},\Delta{R}^2)$ dictates the probability of samples drawn from the fitted Gaussian distribution of observations (Table~\ref{tab:obs}). Additionally, $p(\boldsymbol{\theta}^{(j)}_{\text{EoS}}|\text{samples}^{(j)})=1$ since the reconstruction can locate the deterministic corresponding EoS, given the sampled M-R points, and $p(\text{data}|\boldsymbol{\theta}^{(j)}_{\text{EoS}})\propto \exp{(-\chi^2(M_{\boldsymbol{\theta}^{(j)}_{\text{EoS}}},R_{\boldsymbol{\theta}^{(j)}_{\text{EoS}}})})$ is the likelihood function (of EoS parameters) that estimates the distances of predicted $M-R$'s to the real observations.
%\begin{align}
%    w^{(j)} &= \frac{p(M^{(j)},R^{(j)}|M_\text{obs},R_\text{obs},\theta)}{q(\tilde{M}^{(j)},\tilde{R}^{(j)}|M_\text{obs},R_\text{obs})} \nonumber \\
%      &= \prod_i \frac{p(M_i^{(j)},R_i^{(j)}|M_\text{obs},R_\text{obs}) p(M_i^{(j)},R_i^{(j)}|\theta)}{q(\tilde{M}^{(j)},\tilde{R}^{(j)}|M_\text{obs},R_\text{obs})},
%\end{align}
%where $j$ indicates index of samples and $i$ indicates index of M-R observables. $p(M_i^{(j)},R_i^{(j)}|\theta) = 1$, because it is deterministic. $p(M^{(j)},R^{(j)}|M_\text{obs},R_\text{obs}) \propto \exp{(-\chi^2(M^{(j)},R^{(j)})})$ and $q(\tilde{M}^{(j)},\tilde{R}^{(j)}|M_\text{obs},R_\text{obs}) \sim \mathcal{N}(M(R)_\text{obs},\Delta M(R)^2)$ calculated from the corresponding Gaussian distribution in Table~\ref{tab:obs}. 
In practical calculations, weights should be normalized as $\tilde{w}^{(j)} = w^{(j)} / \sum_j w^{(j)}$ and a cutoff is applied to avoid outliers in the samples. In the training process, target observables of the $\chi^2$ fitting (Eq.~\ref{eq:chi2} in the manuscript) are changed from ($M_{\text{obs},i},R_{\text{obs},i}$) to ($\tilde{M}^{(j)}_i,\tilde{R}^{(j)}_i$). 
%%%%%%%%%%%%%%%%%%%%%%%%%%%%%%%%%%%%%%%%%%%%%%%%%%%%%%%%%%%%
\begin{table}[h!]
\caption{The Gaussian fitted observations based on marginalized distributions of the properties for neutron stars~\cite{Ozel:2015fia,riley:2019nicer}.}
\centering
\def\arraystretch{1.2}
\setlength\tabcolsep{10pt}
\begin{tabular}{@{}lccc@{}}
\hline\hline
Observable & Mass($M_\odot$) & Radius(km)\\
\hline
M13 & 1.42$\pm$0.49  & 11.71$\pm$2.48\\
\hline%
M28 & 1.08$\pm$0.30  & 8.89$\pm$1.16\\
\hline%
M30 & 1.44$\pm$0.48  & 12.04$\pm$2.30\\
\hline%
NGC 6304 &  1.41$\pm$0.54  & 11.75$\pm$3.47\\
\hline%
NGC 6397 &  1.25$\pm$0.39  & 11.48$\pm$1.73\\
\hline%
$\omega$Cen &  1.23$\pm$0.38  & 9.80$\pm$1.76\\
\hline
4U 1608-52 & 1.60$\pm$0.31 & 10.36$\pm$1.98\\
\hline
4U 1724-207 & 1.79$\pm$0.26 & 11.47$\pm$1.53\\
\hline
4U 1820-30 & 1.76$\pm$0.26  & 11.31$\pm$1.75\\
\hline
EXO 1745-248 & 1.59$\pm$0.24 & 10.40$\pm$1.56\\
\hline
KS 1731-260 & 1.59$\pm$0.37 & 10.44$\pm$2.17\\
\hline
SAX J1748.9-2021 & 1.70$\pm$0.30 & 11.25$\pm$1.78\\
\hline
X5  & 1.18$\pm$0.37 &  10.05$\pm$1.16\\
\hline
X7  & 1.37$\pm$0.37 & 10.87$\pm$1.24\\
\hline
4U 1702-429  & 1.90$\pm$0.30 & 12.40$\pm$0.40\\
\hline
PSR J0437–4715  & 1.44$\pm$0.07 & 13.60$\pm$0.85\\
\hline
PSR J0030+0451  & 1.44$\pm$0.15 & 13.02$\pm$1.15\\
\hline
PSR J0740+6620  & 2.08$\pm$0.07 & 13.70$\pm$2.05\\

\hline\hline
\end{tabular}
\label{tab:obs}
\end{table}
%%%%%%%%%%%%%%%%%%%%%%%%%%%%%%%%%%%%%%%%%%%%%%%%%%%%%%%%%%%%

\end{document}